\documentclass[twocolumn,aps,prl,groupedaddress,nofootinbib,showpacs]{revtex4}
\usepackage{graphicx}
\usepackage{amsmath}
\usepackage{amssymb}
\usepackage{multirow}
\usepackage{bm}
\usepackage{color}
\usepackage[hypertex]{hyperref}

\usepackage{relsize}
\RequirePackage{xspace}
\usepackage{dcolumn}
\usepackage{bm}

\def\jpsi{{J/\psi}}
\def\psip{{\psi^\prime}}

\def\ss{{\bigl.^3\hspace{-1mm}S^{[1]}_1}}
\def\sps{{\bigl.^1\hspace{-1mm}S^{[8]}_0}}
\def\so{{\bigl.^3\hspace{-1mm}S^{[8]}_1}}
\def\pjs{{\bigl.^3\hspace{-1mm}P^{[1]}_J}}
\def\pj{{\bigl.^3\hspace{-1mm}P^{[8]}_J}}
\def\p0{{\bigl.^3\hspace{-1mm}P^{[8]}_0}}
\def\sso{\bigl.^3\hspace{-1mm}S^{[1,8]}_1}
\def\sa{{\bigl.^1\hspace{-1mm}S^{[8]}_0}}
\def\sb{{\bigl.^3\hspace{-1mm}S^{[8]}_1}}

\def\be{\begin{equation}}
\def\ee{\end{equation}}
\def\bea{\begin{eqnarray}}
\def\eea{\end{eqnarray}}

\def\gev{\mathrm{~GeV}}

\def\a{\alpha}
\def\th{\theta}

\begin{document}


\title{\mbox{}\\[10pt]
$\jpsi$ polarization at hadron colliders in nonrelativistic QCD}

\author{Kuang-Ta Chao$^{(a,b)}$, Yan-Qing Ma$^{(a,c)}$,Hua-Sheng Shao$^{(a)}$, Kai Wang$^{(a)}$, and Yu-Jie Zhang$^{(d)}$}
\affiliation{ {\footnotesize (a)~Department of Physics and State Key
Laboratory of Nuclear Physics and Technology, Peking University,
 Beijing 100871, China}\\
{\footnotesize (b)~Center for High Energy Physics, Peking
University, Beijing 100871, China}\\
{\footnotesize (c)~Physics Department, Brookhaven National Laboratory, Upton, NY 11973, USA}\\
{\footnotesize (d)~Key Laboratory of Micro-nano
 Measurement-Manipulation and Physics (Ministry of Education) and School of Physics, Beihang University,
 Beijing 100191, China}}




\begin{abstract}
With nonrelativistic QCD factorization, we present a full
next-to-leading order computation of the polarization observable for
$\jpsi$ production at hadron colliders including all important Fock
states, i.e. $\sso,\sps$, and $\pj$.
We find the $\pj$ channel contributes a positive longitudinal
component and a negative transverse component, so the $\jpsi$
polarization puzzle may be understood as the transverse components
canceling between $\sb$ and $\pj$ channels, which results in mainly
the unpolarized (even slightly longitudinally polarized) $\jpsi$.
This may give a possible solution to the long-standing $\jpsi$
polarization puzzle. Predictions for $\jpsi$ polarization at the LHC
are also presented.
\end{abstract}
\maketitle
Nonrelativistic QCD (NRQCD)\cite{Bodwin:1994jh} is an effective
field theory approach for heavy quarkonium. At present, one of the
main obstacles to NRQCD is the polarization puzzle of $\jpsi$
hadroproduction\cite{arXiv:1010.5827}. At leading order (LO) in
$\a_s$, $\jpsi$ production is dominated by gluon fragmentation to a
color-octet (CO) $\sb$ $c\bar c$ pair at high transverse momentum
$p_T$, which leads to transversely polarized
$\jpsi$\cite{hep-ph/9911436}. But the CDF Collaboration found the
prompt $\jpsi$ in its helicity frame to be unpolarized and even
slightly longitudinally polarized\cite{arXiv:0704.0638}. Despite of
numerous attempts made in the past, the puzzle still remains.

For unpolarized $\jpsi$ production, important progress has been made
in recent years. It was found that the next-to-leading order (NLO)
QCD corrections to differential cross sections of $\ss$ channel can
be as large as 2 orders of magnitude at high
$p_T$\cite{Campbell:2007ws}, while those of the $\sa$ and $\sb$
channels are small\cite{Gong:2008ft}. Furthermore, NLO corrections
of the $\pjs$\cite{ma:chic} and
$\pj$\cite{arXiv:1002.3987,Butenschoen:2010rq} channels are found to
also be very large.  These large corrections are well understood
because at NLO the differential cross section
$\rm{d}\sigma/\rm{d}p_T$ receives contributions from new topologies
that scale with $p_T$ in a different manner from the LO calculation.
By including NLO corrections, one may explain the existing
unpolarized cross sections of $p_T$ up to
$70~\rm{GeV}$\cite{ma:chic,arXiv:1002.3987}.

Accordingly, it is necessary to examine the $\jpsi$ polarization at
NLO. Among various channels, the correction to $\jpsi$ polarization
via the $\ss$ channel was worked out in \cite{arXiv:0802.3727},
which alters the polarization from being transverse at LO to
longitudinal at NLO. This phenomenon was explained recently in
collinear factorization~\cite{Kang:2011mg} as next-to-leading power
dominance. As for the $\sb$ channel, the NLO correction can only
slightly change the polarization\cite{Gong:2008ft}, while the $\sa$
channel gives an unpolarized result to all orders in $\a_s$. As a
result, up to date theoretical predictions may indicate a serious
puzzle for $\jpsi$ polarization\cite{Gong:2008ft}. However, the NLO
correction of the $\pj$ channel to $\jpsi$ polarization has not been
calculated.
In this Letter, we perform this calculation and show that the NLO
contribution of $\pj$ channel is indeed crucial in clarifying the
long-standing $\jpsi$ polarization puzzle in NRQCD.

We first introduce some formalisms in our calculation. The $\jpsi$
can decay into an easily identified lepton pair. The information
about the $\jpsi$ polarization is encoded in the angular
distributions of the leptons. The two-body leptonic decay angular
distribution of $\jpsi$ in its rest frame is usually parameterized
as\cite{hep-ph/9709376}
\begin{eqnarray}
\frac{\rm{d}\mathcal{N}}{\rm{d}\cos{\th}}&\propto&
1+\lambda_{\th}\cos^2{\th},\lambda_{\th}=\frac{\rm{d}\sigma_{11}-\rm{d}\sigma_{00}}{\rm{d}\sigma_{11}+\rm{d}\sigma_{00}}.
\end{eqnarray}
Here, $\rm{d}\sigma_{ij}$ ($i,j=0,\pm1$, with respect to the z
components of $\jpsi$) represents the $ij$ contribution in the spin
density matrix formalism.
In the literature, the polarization observable $\lambda_{\th}$ is
also denoted as
$\a=\frac{\rm{d}\sigma_{T}-2\rm{d}\sigma_L}{\rm{d}\sigma_{T}+2\rm{d}\sigma_L}$.
The differential cross sections are
\begin{eqnarray}
\rm{d}\sigma_{s_zs_z}\hspace{-0.1cm}=\hspace{-0.12cm}\sum_{ijn}\hspace{-0.12cm}
{\int\hspace{-0.12cm}{\rm{d}x_1\rm{d}x_2\hspace{-0.02cm}\emph{f}_{i/H_1}
\hspace{-0.05cm}(x_1,\mu_F)
\hspace{-0.03cm}\emph{f}_{j/H_2}\hspace{-0.05cm}(x_2,\mu_F)}}\hspace{-0.03cm}
{\langle\mathcal{O}_n\rangle\rm{d}\hat{\sigma}^{ij,n}_{s_zs_z}},
\end{eqnarray}
where $\langle\mathcal{O}_n\rangle$ are the long-distance matrix
elements (LDMEs) for $n=\sso,\pj,\mathrm{and}~\sps$. In general, the
partonic cross sections $\rm{d}\hat{\sigma}^{ij,n}_{s_zs_z}$ can be
obtained from the spin density matrix elements\cite{hep-ph/9709376}
\begin{eqnarray}
\rho_{s_zs_z}(ij\rightarrow(c\bar{c})[n]X)\propto
\sum_{L_z}|{\mathcal{M}(ij\rightarrow(c\bar{c})[L_z,s_z]X)}|^2.
\end{eqnarray}
In practice, several polarization frame definitions have been used
in the literature. In the s-channel helicity frame, the polar axis
is chosen as the flight direction of the $\jpsi$ in the laboratory
frame. Another frequently used frame is the so-called Collins-Soper
frame\cite{Print-77-0288 (PRINCETON)}.
For simplicity, here we will only choose the helicity frame, the
same as used by CDF\cite{arXiv:0704.0638}. The full theoretical
predictions of azimuthal correlations and the theoretical
descriptions by Collins-Soper, and feed down from $\chi_{cJ}$ and
$\psip$ will be presented in a forthcoming publication.

We describe our method briefly for the sake of completeness. Some
improvements are made in our calculations, while most of our method
has been encompassed in Ref.~\cite{arXiv:1002.3987}. The
calculations of real corrections are based on the Dyson-Schwinger
equations. After absorbing the core codes of the published
HELAC\cite{hep-ph/0002082}, we promote it into a form that can
generate the matrix element of heavy quarkonia (especially P-wave)
production at colliders by adding some P-wave off-shell currents.
The virtual corrections are treated analytically, and the helicity
matrix elements are obtained using the spinor helicity
method\cite{CERN-TH-4186-85}.

\begin{figure}
\includegraphics[width=8.5cm]{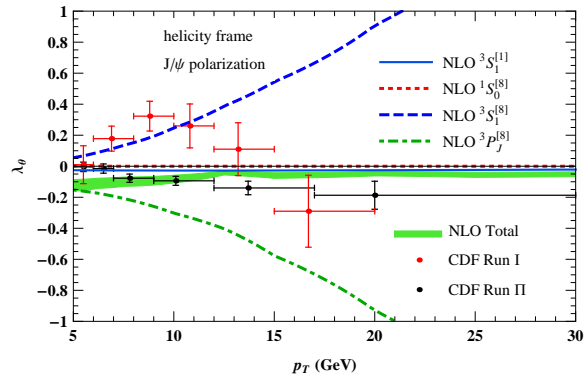}
\caption{\label{fig:polar} (color online) NLO results for the
polarization observable $\lambda_{\th}$ of $\jpsi$ production at the
Tevatron. The CDF experimental data are taken from
Ref.~\cite{arXiv:0704.0638}.}
\end{figure}

For numerical results, we choose the same input parameters as in
Ref.~\cite{arXiv:1002.3987}. Specifically, the renormalization scale
$\mu_r$, factorization scales $\mu_f$ and NRQCD scale
$\mu_{\Lambda}$ are chosen as $\mu_r=\mu_f=m_T=\sqrt{4m_c^2+p_T^2}$
and $\mu_{\Lambda}=m_c$. Scale dependence is estimated by varying
$\mu_r,\mu_f$, by a factor of $\frac{1}{2}$ to $2$ with respect to
their central values. By fitting only the cross sections, it was
found that only two linear combinations of CO LDMEs can be
extracted\cite{arXiv:1002.3987}. Since polarization information is
also available, we will try to extract the three independent CO
LDMEs using the polarization observable $\lambda_{\th}$ and the
production rate $\rm{d}\sigma/dp_{T}$ of the $\jpsi$ measured by CDF
Run II~\cite{arXiv:0704.0638} simultaneously, where the data in the
low transverse momentum region ($p_T<7 \rm{GeV}$) are not included
in our fit because of existing nonperturbative effects. By
minimizing $\chi^2$, the CO LDMEs are obtained and listed in the
first row of Table.\ref{tab:ldmes}. In Fig.\ref{fig:polar}, we
compare $\lambda_{\th}$ from the Tevatron data with our theoretical
results.

\begin{table}[h]
\begin{tabular}{c*{3}{c}}
\hline\hline \itshape ~$\langle\mathcal{O}(\ss)\rangle$~ &
~$\langle\mathcal{O}(\sps)\rangle$~ &
~$\langle\mathcal{O}(\so)\rangle$~ &
~$\langle\mathcal{O}(\p0)\rangle/m_{c}^2$~
\\
\itshape $\rm{GeV}^3$ & $10^{-2}\rm{GeV}^3$ & $10^{-2}\rm{GeV}^3$ &
$10^{-2}\rm{GeV}^3$
\\\hline ~1.16~& $8.9\pm0.98$ & $0.30\pm0.12$ & $0.56\pm0.21$
\\\hline ~1.16~& $0$ & $1.4$ & $2.4$
\\\hline ~1.16~& $11$ & $0$ & $0$
\\\hline\hline
\end{tabular}
\caption{\label{tab:ldmes} Different sets of CO LDMEs for the
$\jpsi$. Values in the first row are obtained by fitting the
differential cross section and polarization of prompt $\jpsi$
simultaneously at the Tevatron~\cite{arXiv:0704.0638}. Values in the
second and third rows are two extreme choices for these CO LDMEs.
The color-singlet LDME is calculated by the B-T potential model in
\cite{Eichten:1995ch}.}
\end{table}

\begin{figure}
\includegraphics[width=7.5cm]{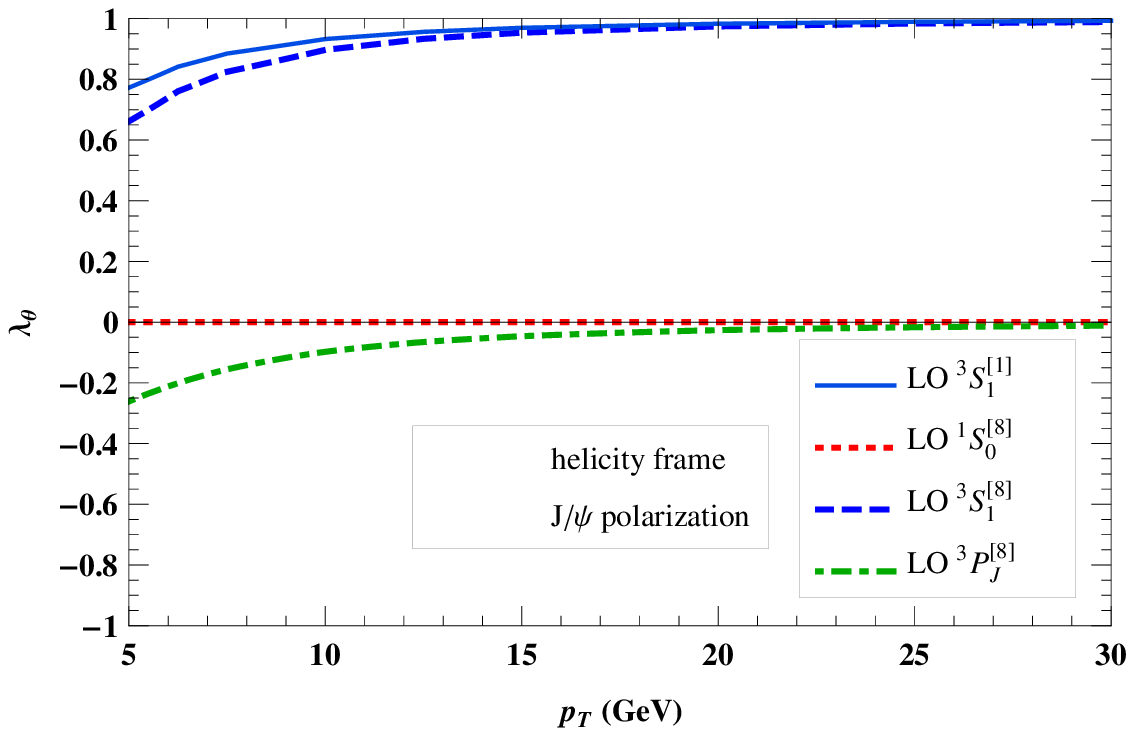}
\includegraphics[width=7.5cm]{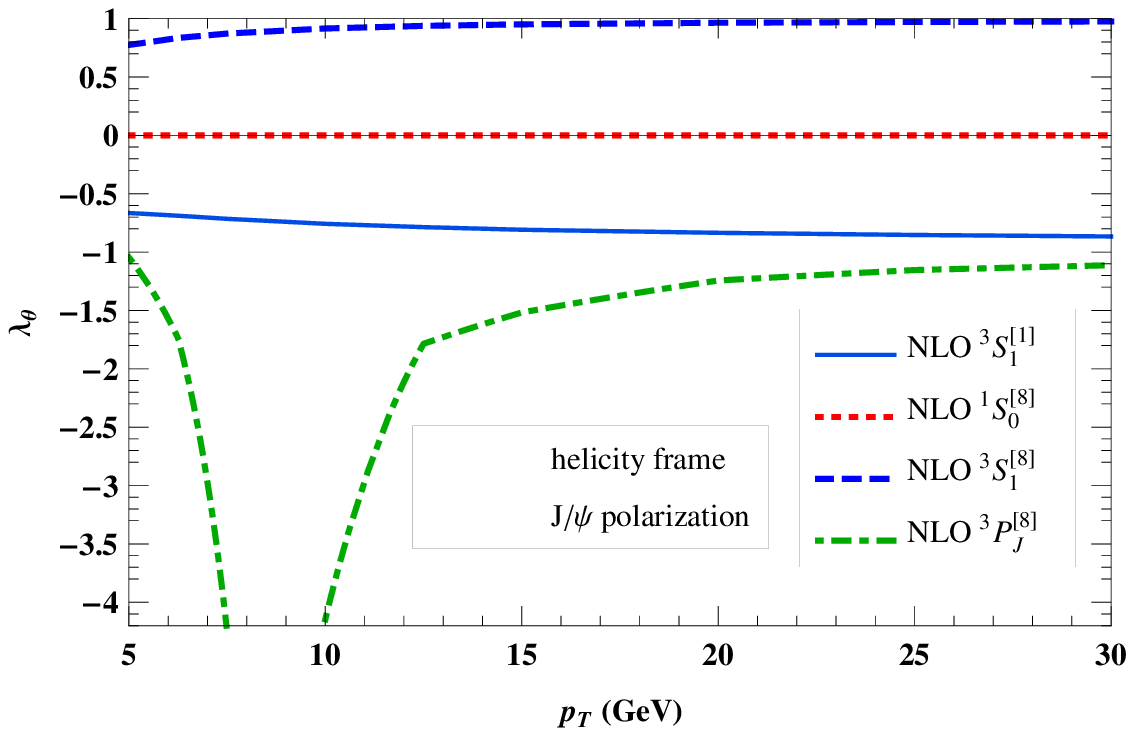}
\caption{\label{fig:partial} (color online) The $p_{T}$ dependence
of $\lambda_{\th}$ for $\ss,\sps,\so,$ and $\pj$ channels with
$\sqrt{S}=1.96~\rm{TeV}$ and $|y_{\jpsi}|<0.6$. For the NLO $\pj$
channel, the dot-dashed curve means the value of
$(\rm{d}\hat{\sigma}_{11}-\rm{d}\hat{\sigma}_{00})/
|\rm{d}\hat{\sigma}_{11}+\rm{d}\hat{\sigma}_{00}| $.}
\end{figure}

To understand the unpolarized results, $\lambda_{\th}$ for each
channel is drawn in Fig.\ref{fig:partial}, where for the NLO $\pj$
channel we mean the value of
$(\rm{d}\hat{\sigma}_{11}-\rm{d}\hat{\sigma}_{00})/|\rm{d}\hat{\sigma}_{11}+\rm{d}\hat{\sigma}_{00}|
$ because $\rm{d}\hat{\sigma}_{11}+\rm{d}\hat{\sigma}_{00}$
decreases from being positive to negative as $p_T$ increases. In
addition to the known polarization of
S-wave\cite{arXiv:0802.3727,Gong:2008ft}, the $\pj$ channel
satisfies $(\rm{d}\hat{\sigma}_{11}-\rm{d}\hat{\sigma}_{00})
/|\rm{d}\hat{\sigma}_{11}+\rm{d}\hat{\sigma}_{00}|<-1$ in our
considered $p_T$ region, which results from
$\rm{d}\hat{\sigma}_{11}<0$ and $\rm{d}\hat{\sigma}_{00}>0$.
Therefore, the transverse component of $\pj$ is negative, which
effectively gives a longitudinal contribution to $\lambda_{\th}$,
and the longitudinal component of $\pj$ is positive. In some
parameter space of the CO LDMEs,
the positive transverse component of  $\sb$  will largely be
canceled by the negative transverse component of $\pj$, which yields
a small transverse component and results in an unpolarized or even
longitudinal $\lambda_{\th}$. This explains why the complete NLO
calculation gives an unpolarized prediction in Fig.\ref{fig:polar}.

It is interesting to see that, by choosing the proper CO LDMEs,
complete NLO predictions in NRQCD factorization can be made
compatible with the data. This is distinct from all previous NRQCD
predictions that give strong transverse polarizations for
$\jpsi$\cite{arXiv:1010.5827}. Furthermore, we want to emphasize the
following four points.

\begin{figure}
\includegraphics[width=8.5cm]{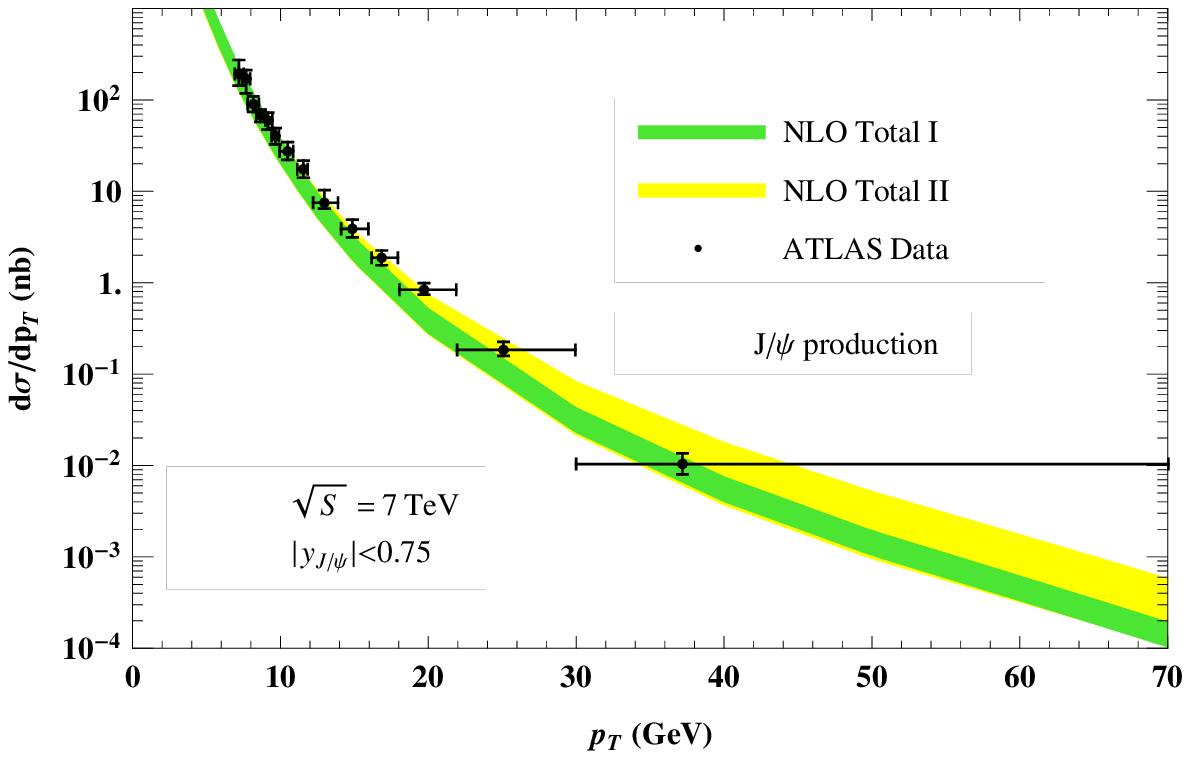}
\includegraphics[width=8.5cm]{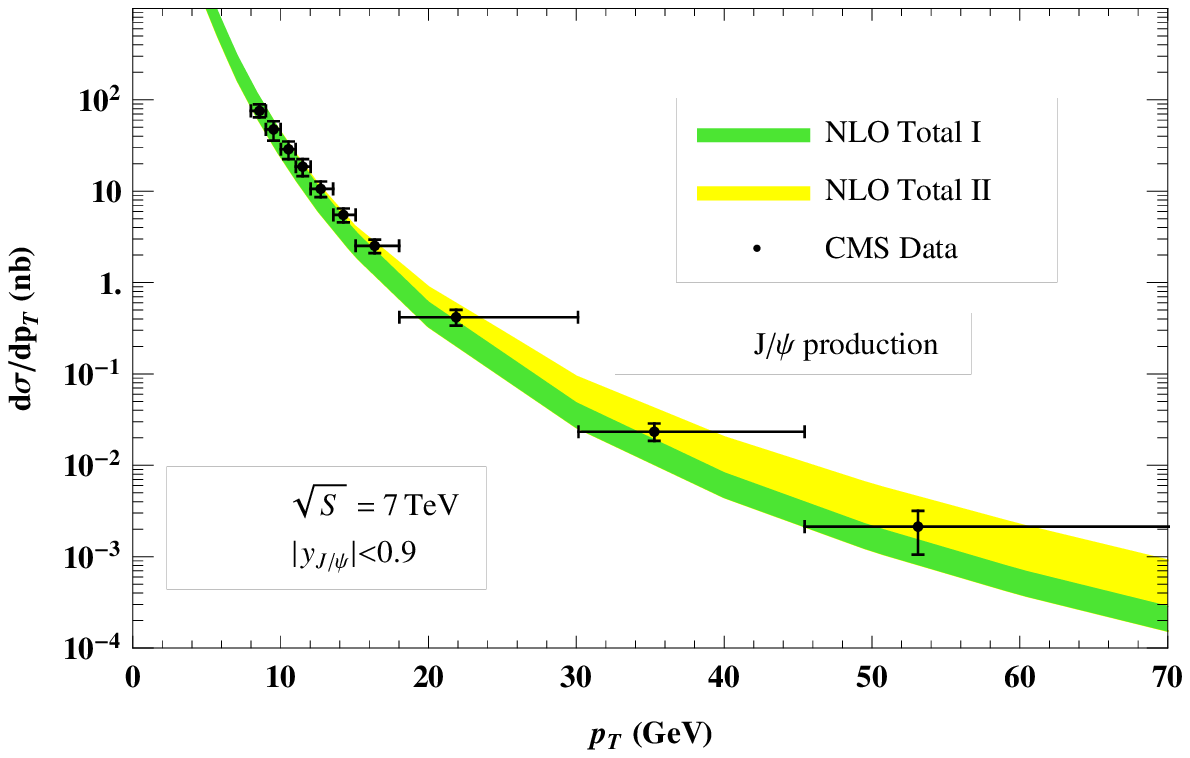}
\caption{\label{fig:cross} (color online) Cross sections of $\jpsi$
production at the LHC with LDMEs shown in Table.\ref{tab:ldmes}.The
green bands (NLO Total I)  correspond to the LDMEs in the first row
of Table.\ref{tab:ldmes}. The yellow bands (NLO Total II) correspond
to the LDMEs in the second row [$\langle\mathcal{O}(\sps) \rangle=0$
(upper bounds)] and third row [$\langle\mathcal{O}(\so)
\rangle=\langle \mathcal{O}(\pj) \rangle=0$ (lower bounds)] of
Table.\ref{tab:ldmes}. The ATLAS data are taken from
Ref.~\cite{ATLASpsi}, and the CMS data from Ref.~\cite{CMSpsi}.}
\end{figure}

(1) Transverse components with large cancellation between $\sb$ and
$\pj$ determine a specific parameter space for the CO LDMEs. Using
the same treatment as in Ref.\cite{arXiv:1002.3987}, we decompose
the transverse component of the short-distance coefficient of $\pj$
into a linear combination of $\so$ and $\sps$ as
$\rm{d}\hat{\sigma}_{11}(\pj)=
2.47~\rm{d}\hat{\sigma}_{11}(\sps)-0.52~\rm{d}\hat{\sigma}_{11}(\so)$.
Since $\rm{d}\hat{\sigma}_{11}(\sps)\ll
\rm{d}\hat{\sigma}_{11}(\so)$ when $p_T >7~\rm{GeV}$, the
cancellation requirement is approximately equivalent to the absence
of the linear combination $\langle \mathcal{O}(\so)
\rangle-0.52~\langle \mathcal{O}(\pj)\rangle/m_c^2$, which is close
to $\rm{M}_1=\langle \mathcal{O}(\so) \rangle-0.56\langle
\mathcal{O}(\pj)\rangle/m_c^2$ defined in
Ref.\cite{arXiv:1002.3987}. Recall that to have a good fit for the
unpolarized yield one needs a very small $\rm{M}_1$, so the
conditions for the CO LDMEs parameter space introduced by fitting
both yield and polarization are consistent with each other. Good
agreement with the LHC data for the $\jpsi$ cross sections can be
found in Fig.\ref{fig:cross} using the LDMEs in
Table.\ref{tab:ldmes}.

(2) As the yield and polarization share a common parameter space,
and the yield can only constrain two linear combinations of CO
LDMEs, the combined fit of both yield and polarization may also not
constrain three independent CO LDMEs stringently. In fact we find
for a wide range of given $\langle\mathcal{O}(\sa)\rangle$, one can
fit both yield and polarization reasonably well. CO LDMEs under two
extreme conditions are listed in Table.\ref{tab:ldmes}. When
$\langle\mathcal{O}(\sa)\rangle$ is chosen to be its maximal value,
the $\jpsi$ is unpolarized; when $\langle\mathcal{O}(\sa)\rangle$
vanishes, $\lambda_\th$ increases from -0.25 at $p_T$=5 GeV to 0 at
$p_T$=15 GeV at the Tevatron. Even in these two extreme cases, the
theoretical predictions of the $\jpsi$ cross section and
polarization are still close to the Tevatron data, and are also
consistent with the observed cross sections obtained by
ATLAS\cite{ATLASpsi} and CMS\cite{CMSpsi} at the LHC as shown in
Fig.\ref{fig:cross}.
As a result, although it is hard to determine the CO LDMEs
precisely, we find that the polarization puzzle can be much eased
for a wide range of $\langle\mathcal{O}(\sa)\rangle$ value.

(3) The cancellation of the transverse component between the $\sb$
and $\pj$ channels is not problematic, since the contribution of an
individual channel is unphysical and depends on renormalization
scheme and scale\cite{arXiv:1002.3987}. A "physical" requirement is
that the summation $\rm{d}\sigma_{11}(\so+\pj)$ be positive, which
is satisfied in the fit.

(4) It is important to note that the LDMEs presented here are
significantly different from those extracted from the global fit in
Ref.\cite{Butenschoen:2011yh}. As hadroproduction data play the most
important role in Ref.\cite{Butenschoen:2011yh}, this difference
cannot be mainly  attributed to data other than hadroproduction not
being considered in our fit. In fact, one can track to the situation
where only hadroproduction data are used in the global fit. As
explained in Ref.\cite{arXiv:1002.3987}, our choice of the $p_T$
cutoff for hadroproduction data is $p_T>7\rm{GeV}$ while the cutoff
in \cite{Butenschoen:2011yh} is $p_T>3\rm{GeV}$, and our LDMEs can
well describe the $p_T$ spectrum in the region
$7\rm{GeV}<p_T<70\rm{GeV}$
(see Fig.\ref{fig:cross}), while the fit in
Ref.\cite{Butenschoen:2011yh} puts stress on the smaller $p_T$
region and gives too smooth a $p_T$ distribution at large $p_T$.
This is the main reason why our LDMEs differ from those in
Ref.\cite{Butenschoen:2011yh}.
In our view, for the small $p_T$ region the fixed order perturbation
calculation may need to be modified by considering soft gluon
emission and other nonperturbative effects. We see that the two
treatments in Ref.\cite{arXiv:1002.3987} and
Ref.\cite{Butenschoen:2011yh} have different features and should be
tested by more experiments in the future.

\begin{figure}
\includegraphics[width=8.5cm]{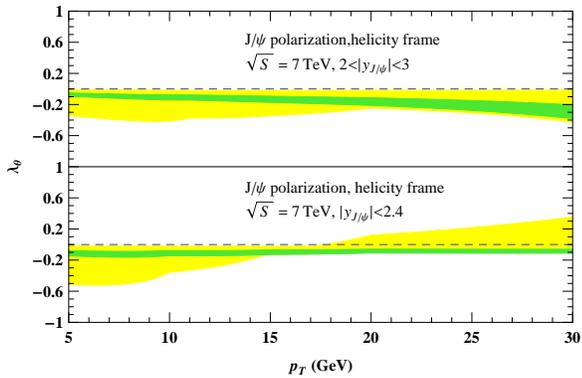}
\caption{\label{fig:polarLHC} (color online) NLO predictions of the
$\jpsi$ polarization observable $\lambda_{\th}$  at the LHC. The
uncertainty is shown by large yellow  bands when varying the CO LDME
$\langle\mathcal{O}(\sa)\rangle$. The bounds of $\lambda_\th=0$ in
yellow bands correspond to CO LDMEs in the third row of
Table.\ref{tab:ldmes}, while the other bounds correspond to the
second row of Table.\ref{tab:ldmes}.  The small green bands are the
predictions using the CO LDMEs in the first row of
Table.\ref{tab:ldmes}.}
\end{figure}

There are still other uncertainties, such as the charm quark mass,
but they do not change the qualitative properties of our result.
Predictions of the polarization  observable $\lambda_{\th}$ at the
LHC with $\sqrt{S}=7~\rm{TeV}$ are plotted in
Fig.\ref{fig:polarLHC}, where only the forward region
($2<|y_{\jpsi}|<3$) and the central region ($|y_{\jpsi}|<2.4$) are
considered\footnote{Note that the ALICE Collaboration has measured
$\jpsi$ polarization recently with rapidity
$2.5<|y_{\jpsi}|<4$\cite{Abelev:2011md}. But the measured transverse
momenta ($2\rm{GeV}<p_T<8\rm{GeV}$) are smaller than considered in
this Letter.}. The large error bar (yellow bands) in these
predictions is caused by a lack of knowledge of
$\langle\mathcal{O}(\sa)\rangle$; thus, we scan all its possible
values in the predictions. It is found in these predictions that
$\lambda_{\th}$ becomes sensitive to
$\langle\mathcal{O}(\sa)\rangle$ when $p_T > 20 \gev$, so it may be
possible to extract three independent CO LDMEs when polarization
data at high $p_T$ are available.


In summary, we present a full NLO calculation including
$\ss,\so,\sps$ and $\pj$ for the polarization observable
$\lambda_{\th}$ of the $\jpsi$ in the helicity frame at the Tevatron
and LHC. Results of S-wave channels are consistent with those in the
literature~\cite{arXiv:0802.3727}, while those of the $\pj$ channel
are new and play a crucial role in understanding the polarization
puzzle. Our calculation shows that the transverse component of the
$\pj$ channel is negative, while its longitudinal component is
positive. Thus the $\pj$ channel gives a maximal longitudinal
contribution. By choosing suitable CO LDMEs, which bring about good
agreement with the observed $\jpsi$ cross sections at large $p_T$ at
the LHC, the transverse components can be largely canceled between
the $\sb$ and $\pj$ channels, leaving the remaining terms to be
dominated by the unpolarized $\jpsi$. This may give a possible
solution to the long-standing $\jpsi$ polarization puzzle within
NRQCD factorization. Although it is hard to individually extract the
three independent CO LDMEs in an accurate way, our interpretation of
$\jpsi$ polarization  makes sense by using only their combinations.
We also present polarization predictions for the LHC.

We thank C. Meng for helpful discussions.  This work was supported
in part by the National Natural Science Foundation of China
(Nos.11021092,11075002,11075011), and the Ministry of Science and
Technology of China (2009CB825200). Y.Q.M is also supported by the
U.S. Department of Energy, contract number DE-AC02-98CH10886.

{\it{Note added.}} When this Letter was being prepared, a
preprint~\cite{BK} on the same issue had just appeared. The
essential difference is that they have a negative
$\langle\mathcal{O}(\p0)\rangle$ based on a global
fit\cite{Butenschoen:2011yh}, and give a significant transverse
polarization prediction, but our fit leads to a positive $\langle
\mathcal{O}(\p0) \rangle$, which is consistent with observed cross
sections in a wide $p_T$ region
(7-70~GeV) at the LHC and results in the mainly unpolarized $\jpsi$.




\end{document}